\documentclass[preprint,1p,12pt]{elsarticle}
\usepackage{amsmath}
\usepackage{amssymb}
\usepackage{placeins}
 \usepackage{setspace}
 \usepackage{natbib}

\journal{Contrast Media \& Molecular Imaging}

\begin{document}

\begin{frontmatter}

\title{Sensitivity Enhancement of (Hyper-)CEST Image Series by Exploiting Redundancies in the Spectral Domain}

 \author{J\"org~D\"opfert}
 \ead{doepfert@fmp-berlin.de}

 \author{Christopher~Witte}
 \ead{witte@fmp-berlin.de}

 \author{Martin~Kunth}
 \ead{kunth@fmp-berlin.de}

 \author{Leif~Schr\"oder\corref{cor1}}
\ead{lschroeder@fmp-berlin.de}

 \cortext[cor1]{Corresponding author}

 \address{Leibniz-Institut f\"ur Molekulare Pharmakologie, Robert-R\"ossle-Str. 10,
    13125 Berlin, Germany}

\begin{abstract}

CEST has proven to be a valuable technique for the detection of hyperpolarized xenon-based functionalized contrast agents. Additional information can be encoded in the spectral dimension, allowing the simultaneous detection of multiple different biosensors. However, due to the low concentration of dissolved xenon in biological tissue, the signal to noise ratio (SNR) of Hyper-CEST data is still a critical issue. In this work, we present two techniques aiming to increase SNR by exploiting the typically high redundancy in spectral CEST image series: PCA-based post-processing and sub-sampled acquisition with low-rank reconstruction. Each of them yields a significant SNR enhancement, demonstrating the feasibility  of the two approaches. While the first method is directly applicable to proton CEST experiments as well, the second one is particularly beneficial when dealing with hyperpolarized nuclei, since it distributes the non-renewable initial polarization more efficiently over the sampling points. The results obtained are a further step towards the detection of xenon biosensors with spectral Hyper-CEST imaging \textit{in vivo}.

\end{abstract}

\begin{keyword}
  hyperpolarized molecules \sep CEST agents \sep low-rank reconstruction \sep undersampling \sep   MRI  \sep NMR \sep  compressed sensing \sep xenon \sep PCA \sep biosensors
\end{keyword}

\end{frontmatter}
\doublespacing

\section{Introduction}
\label{sec_introduction}

Techniques such as hyperpolarization \cite{goodson_nuclear_2002, witte2012} and spin labeling via chemical exchange saturation transfer (CEST) \cite{ward_new_2000, van_zijl_chemical_2011} can be used to overcome the intrinsically low sensitivity of NMR and MRI for detecting molecules of biomedical interest at room temperature. Each method by itself yields a signal enhancement by several orders of magnitude and the combination of both, i.e. Hyper-CEST, has shown great promise for sensitive detection of contrast agents based on functionalized $^{129}$Xe \cite{schroder_molecular_2006-2, kunth_optimized_2012}. 

The large chemical shift range of the noble gas \cite{goodson_nuclear_2002} motivated the design of xenon-based biosensors \cite{spence_functionalized_2001,harel_novel_2008, schroder_xenon_2011} for detecting target molecules and molecular environments of interest \cite{boutin_cell_2011, kotera2012, meldrum_xenon-based_2010} and makes it especially suitable for CEST experiments. Highly sensitive CEST detection of such sensors has been reported, sensing for example incorporation into lipid vesicles \cite{meldrum_xenon-based_2010} and temperature \cite{schroder_temperature-controlled_2008, schroder_temperature_2008,schilling_mri_2010}.
In such setups, xenon atoms reversibly bind to host structures such as cryptophanes \cite{brotin_cryptophanes_2009}, which provide excellent properties for CEST detection. These molecular cages can be functionalized by the attachment of a targeting ligand, thus realizing molecular specificity of the xenon signal.

\begin{figure}[b!]
	 \centering
  \includegraphics[width=0.95\textwidth]{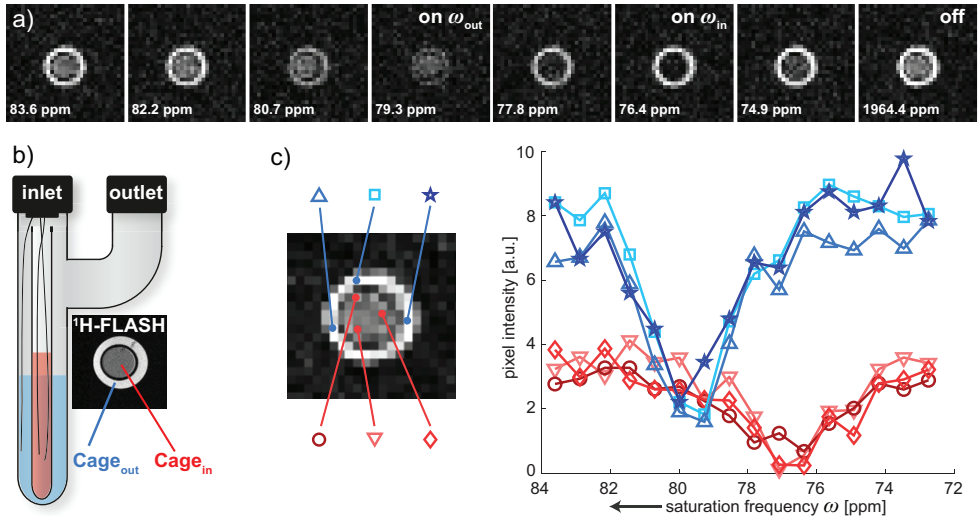} 
  \caption{a) Selected Hyper-CEST images of the two-compartment phantom containing solutions with CEST agents Cage$_\mathrm{in}$ and Cage$_\mathrm{out}$ as depicted in b), acquired with a 32 $\times$ 32 matrix (the complete data set containing all 17 images can be found in the Supporting Information, top row of Fig. S2). The saturation pulse frequency $\omega$ is denoted in the bottom left corner of each image. When $\omega$ approaches the chemical shift of one of the CEST agents ($\omega_\mathrm{in}$ = 76.3~ppm for Cage$_\mathrm{in}$, $\omega_\mathrm{out}$ = 79.5~ppm for Cage$_\mathrm{out}$), the signal of the respective compartment drops due to saturation transfer. b) Schematic of the phantom together with a high-resolution axial slice acquired with proton MRI. c) $\omega$-dependence of the intensity of representative pixels in the outer (red) and the inner compartment (blue). This plot demonstrates that the pixels within each of the two compartments are highly correlated.}
	\label{fig_intro}
\end{figure}

Similar to DIA- and PARACEST experiments, Hyper-CEST \cite{schroder_molecular_2006-2} exploits chemical exchange to transfer labeled spins from a relatively small pool of sensor-associated spins to a more abundant signal, i.e. that of free xenon in solution: A selective radio frequency saturation pulse is used to destroy the polarization of bound xenon, eventually causing an increased signal loss of free xenon in solution which is subsequently imaged. If direct saturation effects are negligible (such as for Hyper-CEST), the normalized CEST effect per pixel can then be determined by subtracting a CEST image acquired with the saturation pulse frequency on-resonant with xenon in the host ($\textbf{S}_\mathrm{on}$) from an off-resonant CEST image ($\textbf{S}_\mathrm{off}$): \textbf{C} $= (\textbf{S}_\mathrm{off} - \textbf{S}_\mathrm{on} ) / \textbf{S}_\mathrm{off}$ \cite{terreno_methods_2009}. Bright regions in the CEST effect map \textbf{C} then indicate the presence of the biosensor. By recording multiple CEST images with different saturation pulse frequencies, it is also possible to encode the chemical shift dimension \cite{kunth_optimized_2012} and hence to acquire spectral CEST image series (see Fig. \ref{fig_intro}a)). This image stack contains a full $z$-spectrum for each pixel (see Fig. \ref{fig_intro}c)), which is particularly useful if the sample contains more than one biosensor or if the exact resonance frequency is unknown due to variations in e.g. temperature or pH.

CEST detection strongly relies on the signal to noise ratio (SNR) of the detected ``abundant'' pool, which is limited for Hyper-CEST by the typically low xenon solubility. This can be partially compensated for by increasing the hyperpolarization or the fraction of the NMR-active isotope $^{129}$Xe.
Here, we present an additional approach based on the inherent redundancy in spectral CEST images. The series of Hyper-CEST images shown in Fig. \ref{fig_intro}a) illustrates this redundancy. We can discern two separate regions, which stem from the two separate compartments in the phantom used throughout this study (Fig. \ref{fig_intro}b)), containing solutions with CEST agents Cage$_\mathrm{in}$  and Cage$_\mathrm{out}$ with different resonance frequencies $\omega_\mathrm{in}$ and $\omega_\mathrm{out}$ (see experimental section for details). Just by visual inspection of the image series, it becomes evident that the data is indeed highly redundant. Each image basically contains the same structure, except that some regions get darker or brighter as the saturation frequency $\omega$ changes. In other words, many pixels vary as a function of $\omega$ in a similar fashion (see also Fig. \ref{fig_intro}c)), for instance the pixels in the outer compartment all decrease in intensity as $\omega$ approaches $\omega_\mathrm{out}$ = 79.5~ppm. The key idea of this work is to exploit these correlations to improve the SNR of such spectral image series. To this end, we propose two compatible methods: The first one filters the noise in the data and is applied as a post-processing step. It is based on principal component analysis (PCA), a commonly used tool to discover and utilize correlations in data \cite{jolliffe_principal_2002}. The second technique utilizes these redundancies directly at the level of image acquisition by recording only a fraction of the conventionally required data points in $k$-space and thereby increasing the signal of Hyper-CEST images. It is related to the first inasmuch as it takes advantage of the very same correlations to correctly reconstruct the full images from the subset of measurements. 

The article presents improved imaging of CEST contrast agents that rely on encoding the spectral dimension. Taking hyperpolarized xenon as an example, it is organized as follows: In the first section, we provide a brief introduction to PCA and then show noise reduction on a fully sampled 32 $\times$ 32 CEST data set (FOV: $2 \times 2$ cm$^2$) by post-processing. The second section initially outlines sub-sampling and constrained reconstruction with the focus on spectral Hyper-CEST images. Finally, we demonstrate the signal gain of such a Hyper-CEST image series on the basis of a sub-sampled 64 $\times$ 64 data set where only one third of the $k$-space lines is acquired.

\section{PCA denoising of spectral CEST images}
\subsection{Background}
\label{sec_backgroundPCA}
PCA is a standard technique to discover and exploit linear correlations in a set of observations \cite{jolliffe_principal_2002}. It extracts the principal components (PCs) --  a set of new, uncorrelated variables -- of the data set and ranks them according to how much of the data's variability they describe. By considering only the highest ranked PCs, i.e. the ones that describe a large portion of the variance of the observations and are hence supposed to convey the essential features of the data,  PCA can be both used for dimensionality reduction \cite{Witte_Extracting_2010} as well as for denoising \cite{balvay_signalnoise_2011}.
We interpret each pixel $(i, j)$ in a stack of spectral CEST images as a variable and the corresponding signal intensity $S_{ij}(\omega_k)$ for a certain saturation frequency $\omega_k$ as an observation of that variable. Assuming that a single CEST image $\textbf{S}(\omega_k)$ is represented as a $u  \times v$ matrix, we can rearrange it as a vector $\textbf{s}(\omega_k)$  with $m = u \cdot v$ entries. We can then reshape the full 3D data set consisting of $n$ images (associated to $n$ different saturation frequencies $\{\omega_1,...,\omega_n\}$) into a $m \times n$ 2D matrix
\begin{equation}
	\textbf{X}=
	\begin{pmatrix}
	& & & \\ \textbf{s}(\omega_1) &  \textbf{s}(\omega_2)  & \ldots & \textbf{s}(\omega_n)\\& & & 
	\end{pmatrix}
	=
	\begin{pmatrix}
	{s}_{1}(\omega_1) &  {s}_{1}(\omega_2)  & \ldots & {s}_{1}(\omega_n)\\
	{s}_{2}(\omega_1) &  {s}_{2}(\omega_2)  & \ldots & {s}_{2}(\omega_n)\\
	\vdots & \vdots & \ddots & \vdots\\
	{s}_{m}(\omega_1) &  {s}_{m}(\omega_2)   & \ldots & {s}_{m}(\omega_n) \\
	\end{pmatrix}
	\label{eq_Xmatrix}
\end{equation}
where the $n$ columns of \textbf{X} contain the pixel observations (i.e. images) for each saturation frequency.
PCA now looks for linear correlations between the pixels with respect to the saturation frequency and yields a matrix \textbf{P} whose rows contain the PCs in descending order. Technically speaking, the PCs are the eigenvectors of the covariance matrix \cite{jolliffe_principal_2002}
\begin{equation}
	\mathrm{cov}(\textbf{X} )=\frac{1}{n-1}\widetilde{\textbf{X}} \widetilde{\textbf{X}}^{\top} 
	\label{eq_cov}
\end{equation}
 ($\widetilde{\textbf{X}}$ denotes the row-wise mean centered version of $\textbf{X}$ and $\widetilde{\textbf{X}}^{\top}$ its transpose), and the associated eigenvalues $\lambda_i$ describe how much of the data's variance is described by each PC. Projecting the original data $\widetilde{\textbf{X}}$ onto the PCs (i.e. by transforming the data into a new basis spanned by the PCs contained in \textbf{P}), one obtains the weightings \textbf{W}:
\begin{equation}
	\textbf{W}=\textbf{P}\widetilde{\textbf{X}} \quad.
\end{equation}
The original mean centered data can now be exactly reproduced using all PCs
\begin{equation}
	\widetilde{\textbf{X}}=\textbf{P}^\top \textbf{W} \quad,
\end{equation}
or it can be approximately reconstructed using only the $r$ highest ranked principal components
\begin{equation}
\widetilde{\textbf{X}}_r=\textbf{P}^\top_r \textbf{W}_r
\label{eq_pca_approxreco}
\end{equation}
where $\textbf{P}_r$ and $\textbf{W}_r$ contain only the first $r$ rows of \textbf{P} and \textbf{W}.

\subsection{Results  \& Discussion}

\begin{figure}[t!]
	 \centering
  \includegraphics[width=0.9\textwidth]{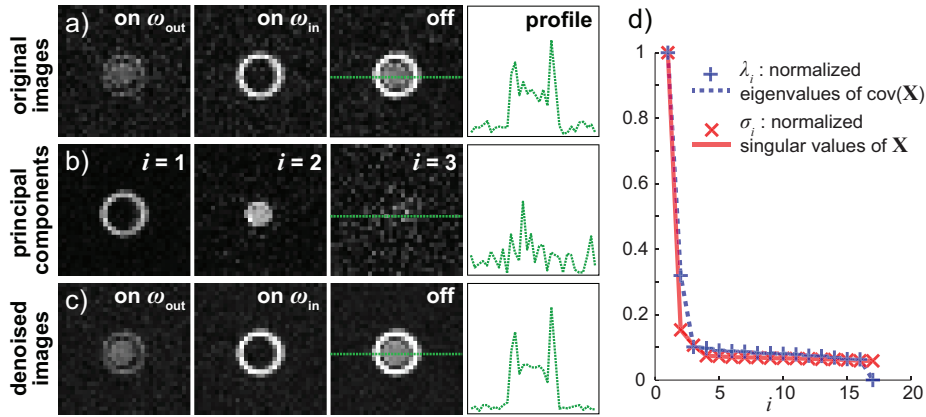} 
   \caption{a) Hyper-CEST images from the same data set as in Fig. \ref{fig_intro}a) acquired with a 32 $\times$ 32 matrix, including a profile plot for better visualization of the noise characteristics. From the full series, only the on- and off-resonant images are shown. b) First three principal components of the data set (absolute values). All but the two highest ranked PCs mostly describe noise. c) The same selection as in a), but this time reconstructed using only the first two PCs (the full data set is shown in the Supporting Information, Fig. S2). Compared to a), the noise is reduced by PCA reconstruction, increasing the SNR of the off-resonant image from $8.5 \pm 0.4$ to $18.7 \pm 0.3$, which corresponds to a factor of $2.2 \pm 0.1$ (the off-resonant image was chosen for SNR determination because its signal is not dependent on any saturation transfer effect). d) Normalized eigenvalues of the covariance matrix of the data set \textbf{X} (blue $+$) and singular values of \textbf{X} (red $\times$). Since each eigenvalue $\lambda_i$ represents how much of the data's variance is described by the $i$th PC, this plot indicates that the first two PCs describe a large portion of the variability of the data. Similarly, the distribution of singular values $\sigma_i$ suggests that the data can be approximated by a low-rank matrix.}

	\label{fig_resPCA}
\end{figure}

In Fig. \ref{fig_resPCA}a), we have reproduced the two on-resonant and the off-resonant images from the fully sampled 32 $\times$ 32 CEST data set of Fig. \ref{fig_intro}b) to facilitate easy comparison. 
As described in section \ref{sec_backgroundPCA}, PCA exploits the inherent pixel correlations to deconstruct the data set into its principal components (see Fig. \ref{fig_resPCA}b)) and ranks them according to how much of the data's variance they describe (see Fig. \ref{fig_resPCA}d), blue $+$). We expect the two highest ranked PCs to describe most of the relevant information, since the phantom contains CEST agents associated to two different frequencies in the xenon spectrum (note that due to mean centering, regions with constant intensity that are not affected by the saturation pulse at all would not be discernible in the PCs but contribute only noise). Fig. \ref{fig_resPCA}b) and \ref{fig_resPCA}d) indeed confirm this consideration; all the lower ranked components mostly contain noise (this was also verified by observing different reconstructions of the data set with a varying number of PCs, see Supporting Information, Fig. S2). Therefore, we reproduce the entire data set using only the two highest ranked PCs (as described by Eq. (\ref{eq_pca_approxreco})) aiming to reduce the noise. The resulting reconstruction in Fig \ref{fig_resPCA}c) shows that PCA can indeed remove much of the noise while the relevant information is preserved, yielding an SNR increase of a factor of $2.2 \pm 0.1$ compared to the original data set. Note that since each compartment contains only one CEST agent, the PCs (see Fig. \ref{fig_resPCA}b)) illustrate the isolated CEST responses in an optimal way without being affected by spectral overlap as e.g. in Fig. \ref{fig_resPCA}c) left, where not only the outer compartment is saturated, but also partly the inner one.

Since the CEST effect of the agents is finally obtained by subtracting each of the two on-resonant images from the off-resonant image, it can be argued that it would suffice to acquire just the three mentioned images instead of the whole spectral series. Investing the same amount of time as for the spectral data set containing 17 images, one could in fact average each of these three images 17/3 times, yielding an SNR increase of a factor of $\sqrt{17/3} = 2.4$ and thereby outperforming the proposed PCA-based reconstruction. 
However, since often the exact frequencies of the agents are not known a priori, multiple images have to be acquired over a certain spectral range anyway. Instead of discarding most of them by calculating the CEST effect the usual way, PCA is able to utilize the information of all images to decrease the noise. 
Furthermore, the shape of the $z$-spectrum bears additional information \cite{zaiss_analytical_2012}, which cannot be resolved by just acquiring the on and off-resonant images. 
Moreover, when the CEST resonances of the agents overlap, then four instead of three images are needed to correctly determine the CEST effect of each agent (see Supporting Information, Fig. S5), diminishing the mentioned SNR advantage of conventional averaging. 
Another technique exploiting spectral information proposed earlier in this journal \cite{terreno_methods_2009} achieves an enhancement of the CEST effect by integrating the $z$-spectrum over several frequencies. In contrast to PCA, which always utilizes all images in the data set, this method becomes less efficient if multiple CEST resonances are present that are very close to each other since then only a few points can be included into the integration. However, PCA denoising and integral-based determination of the CEST effect could be readily combined.

These points suggest that PCA-based reconstruction of CEST data is valuable in various cases. The applicability of the method is not restricted to hyperpolarized nuclei or phantom experiments with a well-defined geometry: As long as there is no object motion/deformation (or it has been corrected), $z$-spectral images of any structure are expected to be highly redundant, since regions containing the same CEST agent typically vary in a correlated manner as a function of saturation frequency. Therefore, the proposed method can also be directly applied to conventional proton CEST imaging.

It should be noted that in biological samples or \textit{in vivo}, parameters such as pH, temperature and agent concentration might vary from voxel to voxel, causing less correlation in between the image pixels and hence additional significant principal components. However, we assume that as long as the maximum CEST effect of a certain agent occurs at the same frequency for each pixel (which is mostly true if $B_0$ inhomogeneities have been corrected), the pixel-wise variations of the CEST response of this agent can still be described by a combination of a small number of principal components.

If most of the data that is recorded within a $z$-spectrum is actually redundant, it seems natural to ask if it is at all necessary to collect all of the data. The main aspect of the next section is therefore to further utilize these redundancies at the earlier stage of image acquisition by sub-sampling $k$-space.

\section{Reconstruction of sub-sampled spectral Hyper-CEST images}
\subsection{Background}
Acquiring less $k$-space data than required by conventional sampling theory, commonly referred to as sub- or undersampling, has been successfully applied in a variety of proton MRI applications. The reduced amount of data leads to a reduced scan time, which enables for example higher temporal resolution in dynamic imaging \cite{tsao_unifying_2001, pedersen_2009, lingala_accelerated_2011, gao_compressed_2012} or, as recently shown, accelerated diffusion-weighted \cite{adluru_g._improving_2012} and CEST MRI \cite{varma_keyhole_2012}. In our setup however, the re-delivery of fresh hyperpolarized noble gas and the subsequent saturation period make up the major fraction ($> 93\%$) of the total scan time, rendering such a speedup of the signal acquisition process less useful. However, sub-sampling can bear another advantage when hyperpolarized nuclei are involved: 
In contrast to thermally polarized samples, the magnetization $M$ does not relax back to its initial, hyperpolarized state $M_0$ once it has been excited into the transverse plane. Hence, every $k$-space line that is acquired within the gradient echo readout uses up a certain fraction of  $M_0$.  
With this in mind, the variable flip angle technique introduced in ref. \cite{zhao_gradientecho_1996} that we employed ensures that the same amount of transverse magnetization (i.e. signal) is available for each of the recorded lines. It is evident that the fewer lines are acquired, the higher the fraction of magnetization that is available to encode each line. One of the key ideas of this article is therefore that sub-sampling can actually increase the SNR of spectral Hyper-CEST images. 
However, if sparsely sampled raw data is reconstructed the usual way by Fourier transform (FT), artefacts occur (see Fig. \ref{fig_resCR}a)) due to violation of the Nyquist criterion. Therefore, more sophisticated reconstruction strategies incorporating prior knowledge about the object have to be applied. 
Techniques that are based on exploiting signal sparsity or compressibility in a known transform domain (compressed sensing \cite{lustig_sparse_2007}) are an active area of research. Here we adopt another but somewhat similar approach: Recently, in the field of matrix completion, it has been shown that low-rank matrices can be recovered from a small subset of random entries \cite{candes_exact_2009,candes_matrix_2010}. A matrix \textbf{X} possesses approximately low rank if a small number of its singular values are much larger than all the others. It is well known that the singular values $\sigma_i$ of \textbf{X} are strongly related to the eigenvalues $\lambda_i$ of the covariance matrix of \textbf{X} (the $\sigma_i$ are the square root of the eigenvalues of $\textbf{X}\textbf{X}^\top$ and the $\lambda_i$ are the eigenvalues of  $\widetilde{\textbf{X}}\widetilde{\textbf{X}}^\top$ up to a normalization factor, see Eq. (\ref{eq_cov})). This implies that a data set that can be mostly described by a few principal components also exhibits a small number of large singular values. As shown in Fig. \ref{fig_resPCA}d), this indeed precisely applies to the matrix \textbf{X} containing the CEST series and hence we decided to recover our sub-sampled data assuming low rank as prior knowledge, similar to ref. \cite{adluru_g._improving_2012}. To do so, we used the optimization algorithm developed by Cai et. al. \cite{cai_singular_2010}, which, broadly speaking, among the infinite number of possible solution matrices \textbf{X} that match the acquired data, picks the one that has minimum nuclear norm (the nuclear norm is the tightest convex relaxation of matrix rank, just as $l_1$ minimization is the the tightest convex relaxation of the $l_0$ minimization problem known from compressed sensing \cite{candes_matrix_2010}). 

\subsection{Results  \& Discussion}
\begin{figure}
	 \centering
  \includegraphics[width=\textwidth]{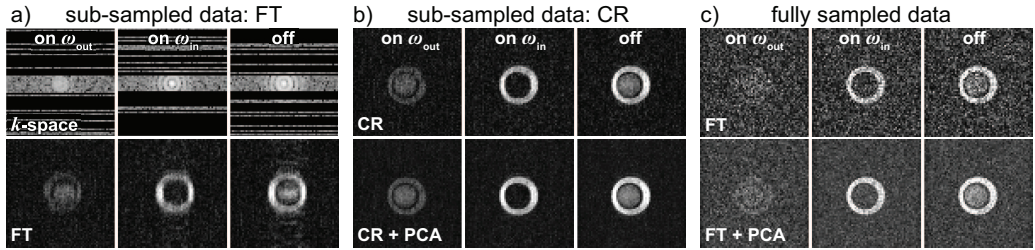}
  \caption{Representative sub-sampled and fully sampled data sets acquired with a 64 $\times$ 64 matrix. From the full series (see Supporting Information, S3), only the on- and off-resonant images are shown. a) Upper row: $k$-space for the sub-sampled data set, where only 34\% of the 64 lines were acquired in a pseudo-random fashion, as described in the experimental section. The missing lines are zero-filled. The corresponding FT-reconstruction is shown in the lower row, exhibiting artefacts along the direction of the sub-sampling. b) Constrained reconstruction (CR) of the sub-sampled data set, without and with PCA post-processing (SNR: $7.8 \pm 0.2$ and $15.3 \pm 0.2$, respectively). CR successfully managed to remove the artefacts present in a). c) The corresponding fully sampled data set without and with PCA post-processing (SNR: $3.1 \pm 0.1$ and $6.8 \pm 0.8$, respectively) exhibits much lower signal. In fact, by using sub-sampling together with PCA (lower row in b)), SNR is increased by a factor of $4.9 \pm 0.1$ compared to the plain FT-reconstruction in the upper row of c).  All SNR values were calculated from the off-resonant image.}
	\label{fig_resCR}
\end{figure}

The upper row of Fig. \ref{fig_resCR}c) shows selected images from a Hyper-CEST data set acquired with the same parameters as the one in Fig. \ref{fig_resPCA}a), but with a matrix size of 64 $\times$ 64 instead of 32 $\times$ 32. 
Due to the higher resolution, the SNR decreases considerably. Sampling only 34 \% of $k$-space as shown in Fig. \ref{fig_resCR}a) distributes a higher fraction of magnetization among each phase encoding line, which clearly increases the signal compared to c), but also leads to artefacts in the direction of undersampling when conventional Fourier reconstruction is used for image recovery. These artefacts can be efficiently removed using a rank constrained reconstruction (CR), as displayed in the upper row of Fig. \ref{fig_resCR}b). Here, the SNR is increased by a factor of $2.5 \pm 0.1$ compared to the FT reconstruction of the fully sampled data set in c). In fact, the SNR is almost as high as in the fully sampled 32 $\times$ 32 data set (Fig. \ref{fig_resPCA}a), demonstrating that the proposed method enables a 4-fold increase in resolution (2-fold in each dimension) with only a marginal cost in image quality ($\mathrm{SNR}_{64\mathrm{sub}}=92\% \cdot \mathrm{SNR}_{32\mathrm{full}}$). 

We observed that the CR-recovered image series still had a full matrix rank of 17, even though the reconstruction procedure is expected to minimize the nuclear norm and thereby the rank. This is most probably due to the presence of noise, which causes the spectral CEST data to be not exactly but only approximately low-rank (this is reflected in the non-zero eigenvalues $\lambda_i$ with $i>2$ associated to the noisy principal components in Fig. \ref{fig_resPCA}d)). Since the reconstruction enforces consistency with the actually measured $k$-space lines, the encountered solution still exhibits a considerable amount of uncorrelated noise. Recalling the results from the previous section, it is reasonable to assume that PCA can remove much of these random fluctuations. Therefore, we further post-processed the sub-sampled and CR-recovered data set with the PCA-based method presented in the previous section. This raised SNR further by a factor of $2.0 \pm 0.1$ (Fig. \ref{fig_resCR}b), lower row), thus being now as much as $4.9 \pm 0.1$ times higher than the SNR for the original fully sampled data set without post-processing (Fig. \ref{fig_resCR}c), upper row). Once again, note that sub-sampling yields mostly signal gain whereas PCA yields noise removal. Therefore, the PCA-processed sub-sampled data set (lower row in Fig. \ref{fig_resCR}b)) still achieves much higher SNR than the PCA-processed fully sampled data set (lower row in Fig. \ref{fig_resCR}c)).
Just as with PCA, the rank-minimizing reconstruction employed here does not make any assumptions about the structure of the object, since it only exploits correlations in the $\omega$-dimension and can hence be applied to any spectral CEST data set. However, one could probably further increase the sub-sampling (i.e. reduce the amount of acquired data) by imposing additional prior knowledge on the images such as sparsity in a known transform domain (e.g. wavelet or total variation) together with rank regularization as recently proposed  \cite{lingala_accelerated_2011,gao_compressed_2012}. 

The theoretical results that guarantee exact matrix recovery by rank minimization have been derived assuming uniform random sub-sampling \cite{candes_matrix_2010}. However, the sampling of random points in $k$-space is impossible to achieve with the conventional Cartesian gradient echo readout used here, since a complete $k$-space line is acquired for each excitation. Hence, the use of more irregular sampling strategies such as radial \cite{lingala_accelerated_2011,ktblock2007} or spiral \cite{santos_2006} trajectories that are `closer' to random sampling might enable a better reconstruction.
In general, we found that the acquisition of more data (i.e. bigger matrix size, more saturation offsets $\omega$) allows for higher undersampling factors, probably due to the fact that most of the additionally acquired pixels or images within a CEST data set are actually increasing its redundancy. 
Furthermore, the sampling gets more incoherent, since the more data points a full data set consists of, the better it is possible to distribute the actually acquired sampling points in a pseudo random fashion. Therefore, undersampling on 32 $\times$ 32 datasets is less beneficial than on 64 $\times$ 64 datasets.

For sufficiently long $T_2^*$ relaxation times, single shot sequences like echo planar imaging (EPI) may provide better SNR for hyperpolarized xenon imaging \cite{kunth_optimized_2012} than the gradient echo sequence (GRE), since the whole hyperpolarized magnetization $M_0$ is used to encode the full image and not shared among multiple $k$-space lines. However, we chose GRE here for two reasons: First, in contrast to EPI, GRE is much less prone to chemical shift artefacts. Second, the application of a variable flip angle, which keeps the fraction of magnetization available for encoding each phase encoding line equal, ensures that the signal strength directly depends on the sub-sampling factor (see Eq. (\ref{eq_vfa_signal})). In an EPI sequence for example, subsequent data points suffer from signal loss due to $T_2^*$ relaxation, which is why high readout bandwidths are usually used to guarantee enough signal also for the final phase encoding lines. Although the use of undersampling here would still enable SNR improvement (the bandwidth could be lowered), it would be not as well-defined as in the case of GRE imaging. Note that a very fast $T_2^*$-decay of the xenon signal, e.g. when it is bound to cell membranes, can make it difficult to use EPI.

Both the variable flip angle approach and EPI require that a single xenon delivery provides enough initial magnetization that can be distributed among multiple $k$-space lines. If this condition is not fulfilled, sub-sampling may only help to reduce the number of repetitions. Using strong laser systems for the hyperpolarization, it has been demonstrated that sufficiently strong magnetization can be provided \cite{kunth_optimized_2012}.

While the results derived in the previous section regarding PCA-based noise reduction are directly applicable to conventional proton CEST MRI, this is not the case for sub-sampling: Since then, due to $T_1$-recovery of $M_0$, the signal available for each $k$-space line in the steady state does not depend on the sub-sampling factor, acquiring fewer data yields an acceleration of encoding time \cite{varma_keyhole_2012} instead of a signal gain. Nevertheless, the saved time increases the CNR efficiency (contrast-to-noise ratio divided by the square root of the acquisition time) \cite{liu09} and might be invested in signal averaging. 

To summarize this section, we showed that sub-sampling can indeed significantly increase the signal when applied to spectral Hyper-CEST data. Using a GRE readout combined with a variable flip angle, it allowed for a 4-fold increase in resolution with almost no cost in image quality.

\section{Summary \& Conclusion}
We demonstrated the presence of redundancies in typical spectral CEST image series and proposed two methods to exploit these correlations for improving image SNR: PCA-based post-processing and sub-sampled acquisition of CEST data sets.  
First, using PCA-enabled denoising, we were able to more than double the SNR of a spectral CEST series at a matrix size of $32 \times  32$. 
Second, the signal gain per $k$-space line achieved by sub-sampled acquisition with a variable flip angle followed by constrained image reconstruction allowed for a 2.5-fold increase in SNR of a spectral Hyper-CEST series at a matrix size of $64 \times  64$ compared to conventional sampling. Therefore, sub-sampling enabled a 4-fold increase in resolution at almost no cost in image quality compared to fully sampled lower resolution images.
Since the reconstructed, sub-sampled data sets still exhibited uncorrelated noise, we additionally applied post-processing with PCA, increasing SNR by another factor of 2.
These findings suggest that the clever utilization of redundancies in spectral CEST images can be very beneficial. Especially when dealing with contrast agents responsive at various frequencies such as pH sensitive CEST agents or functionalized xenon biosensors based on hyperpolarized nuclei, the proposed methods provide increased image quality without the need for time-consuming signal averaging. Therefore, they might be of great use for CEST imaging in general and for future spectral Hyper-CEST imaging experiments \textit{in vivo} in particular.

\section{Experimental}

\subsection{Setup}

All MRI experiments were performed at room temperature ($\approx $ 295~K) on a 9.4~T NMR spectrometer (Bruker Biospin, Ettlingen, Germany) equipped with gradient coils for imaging. A 10~mm inner diameter double-resonant probe, tuneable to both $^{129}$Xe and $^1$H, was used for excitation and detection. We controlled the gas flow by controllers at the sample gas outlet. Hyperpolarized  $^{129}$Xe was generated by spin exchange optical pumping (ca. 16\% spin polarization after transfer into the NMR spectrometer) in a custom designed continuous flow setup using a gas mixture of 5\% xenon (26.4\% natural abundance of  $^{129}$Xe), 10\% N$_2$ and 85\% He. Using the pressure from the polarizer (ca. 4.5 atm. abs.), the mix was directly bubbled into the sample solution \cite{witte_c._hyperpolarized_2012} for 11~s at a total flow rate of 0.1~SLM (standard liter per minute) followed by an 0.5~s delay before signal encoding (to allow the remaining bubbles to collapse). 
Measurements were conducted on a phantom consisting of two compartments (see Fig. 1b)): The inner compartment contained 80\% vol. DMSO and 20\% vol. H$_2$O, and the outer compartment contained 95\% vol. DMSO and 5\% vol. H$_2$O. In both compartments, Cryptophane-A mono-acid cages were dissolved at a concentration of 80~$\mu$M. The different DMSO content yielded a chemical shift separation of the cage resonances for the inner and the outer compartment, mimicking two different CEST agents \cite{kunth_optimized_2012}: Cage$_\mathrm{in}$  at $\omega_\mathrm{in} =$ 76.3~ppm and Cage$_\mathrm{out}$ at $\omega_\mathrm{out} =$79.5~ppm (see the direct NMR spectrum in Supporting Information, Fig. S1). Since also the xenon solubility depends on the DMSO content, the MR signal intensity of the outer compartment was higher than the signal intensity of the inner compartment in all figures of this article.

\subsection{Hyper-CEST MRI}

All $^{129}$Xe CEST images were acquired using a standard gradient echo (GRE) readout (20 $\times$ 20~mm field of view, 20~mm slice thickness, a bandwidth of 12~kHz, an echo time of 5.3~ms and a repetition time of 16~ms) with a slice-selective Gaussian-shaped excitation pulse. 

A variable excitation flip angle $\alpha_i$, $i = \{1,...,n\}$, was used to ensure that the same amount of initial transverse magnetization was available for the encoding of each of the $n$ lines in $k$-space \cite{zhao_gradientecho_1996}:
\begin{equation}
\alpha_i=\arctan\left(\frac{1}{\sqrt{n-i}} \right) \quad .
\label{eq_vfa}
\end{equation}
Since the amount of transverse magnetization is the same for each line, it can be calculated using the first flip angle:
\begin{equation}
	M_{xy}=M_0 \sin(\alpha_1)=\frac{M_0}{\sqrt{n}}
	\label{eq_vfa_signal}
\end{equation}
where $M_0$ denotes the initial longitudinal magnetization and the identity $\sin(\arctan(x))=x/\sqrt{1+x^2}$ was used.

To achieve CEST weighting, prior to each image acquisition a continuous wave saturation pulse of amplitude $B_1$ = 5~$\mu$T and 2~s duration was applied at a frequency  $\omega_i$. For each CEST series, 17 CEST images were acquired: 16 with $\omega_i$ being swept in equidistant steps from 83.6~ppm to 72.8~ppm (to cover both the Cage$_\mathrm{in}$ and the Cage$_\mathrm{out}$  resonances), and one off-resonant image with  $\omega_i$ = 1964.4~ppm (with respect to the frequency of xenon in gas phase). 

We acquired three data sets: Two fully sampled series with a matrix size of 32 $\times$ 32 and 64 $\times$ 64 (GRE readout times 512 and 1024~ms, respectively), and one sparsely sampled series with a matrix size of 64 $\times$ 64, where only 22 out of 64 lines (34\% of the data) were actually acquired (GRE readout time: 352~ms). Since most of the signal energy (and the main contrast information) of natural images is expected to be stored in the $k$-space center \cite{lustig_sparse_2007}, 9 of the 22 lines were used to cover the central lines, whereas the remaining 13 lines were acquired uniformly at random for each $\omega_i$ (see Supporting Information, Fig. S4 for further details). To improve comparability, the repetition time, the bandwidth and the echo time were kept constant for both the 32 $\times$ 32 and the 64 $\times$ 64 GRE readouts and not set to their minimum values.
All measurements were repeated three times to enable error estimation.

\subsection{Post-Processing}

All post-processing routines were written in Matlab (R2012a). PCA reconstruction was run on the magnitude images of a CEST series and implemented with singular value decomposition. The two highest ranked PCs were then used for PCA reconstruction. 

Constrained reconstruction of sparsely sampled data using nuclear norm minimization was performed with a singular value thresholding algorithm \cite{cai_singular_2010} adapting the code from Cand\`{e}s et al. \cite{candes_code}. Reconstruction of the sub-sampled CEST datasets took on average 25 s on a standard desktop computer.  For more details about the reconstruction parameters, see the Supporting Information, section 3.2. 

SNR was calculated by placing regions of interest (ROIs) into the regions containing signal (inner and outer compartment of the phantom) and the regions containing noise. Then, the mean pixel grey value of the signal region $\mu_\mathrm{signal}$ was divided by the standard deviation of the pixel values of the noise region $\sigma_\mathrm{noise}$ and corrected by a factor of $\sqrt{2-\pi/2}$ for Rician noise \cite{gudbjartsson_rician_1995}:
\begin{equation}
	\mathrm{SNR}=\frac{\mu_\mathrm{signal}}{\sigma_\mathrm{noise}} \sqrt{2-\frac{\pi}{2}} \quad.
\end{equation}
All SNR values in the text and their errors are obtained calculating mean and standard deviation of the three SNR values based on the three repetitive acquisitions of each data set.

\section{Acknowledgements}
The authors would like to thank Matthias Schnurr for valuable comments and suggestions.
This work has been supported by the European Research Council
under the European Community's Seventh Framework Programme
(FP7/2007-2013)/ERC grant agreement no. 242710 and the Human
Frontier Science Program.

\bibliographystyle{bib_style}
\bibliography{my_bib_abbr}

\end{document}